\documentclass[11pt]{amsart}
\usepackage{latexsym,amssymb,amsmath,amscd,amsthm}
\numberwithin{equation}{section}
\theoremstyle{remark}

\newcommand{\bq}{\begin{equation}}
\newcommand{\bea}{\begin{array}}
\newcommand{\eea}{\end{array}}

\newcommand{\ga}{\alpha}

\newcommand{\gD}{\Delta}

\newcommand{\gb}{\beta}

\newcommand{\mf}{\mathfrak}
\newcommand{\mc}{\mathcal}

\newcommand{\gO}{\Omega}

\newcommand{\gd}{\delta}
\newcommand{\pp}{\partial}

\newcommand{\tl}{\tilde}
\newcommand{\na}{\nabla}

\newcommand{\bs}{\blacksquare}

\newcommand{{\DDD}}{D\!\!\!\!\!\!-}

\newcommand{\bx}{\Box}

\title{SOME REMARKS ON RICCI FLOW AND THE QUANTUM POTENTIAL}

\author{Robert Carroll\\University of Illinois, Urbana, IL 61801}

\date{March, 2007\thanks{email: rcarroll@math.uiuc.edu}}

\begin{document}

\bibliographystyle{plain}

\begin{abstract} 
We indicate some formulas connecting Ricci flow and Perelman entropy to 
Fisher information, differential entropy, and the quantum potential.
\end{abstract}

\maketitle


\section{FORMULAS INVOLVING RICCI FLOW}
\renewcommand{\theequation}{1.\arabic{equation}}
\setcounter{equation}{0}

Certain aspects of Perelman's work on the Poincar\'e conjecture have applications
in physics and we want to suggest a few formulas in this direction; a fuller 
exposition will appear in a book in preparation
\cite{cczz}.  We go first to \cite{chkf,jujt,khlo,khbl,klfd,mulr,perl,topp} and simply write
down a few formulas from \cite{mulr,topp} here with minimal explanation.
Thus one has Perelman's functional ($\dot{{\mc R}}$ is the Riemannian Ricci
curvature)
\bq\label{1.1}
{\mf F}=\int_M(\dot{{\mc R}}+|\na f|^2)exp(-f)dV
\end{equation}
and a so-called Nash entropy $({\bf 1A})\,\,N(u)=\int_Mulog(u)dV$ where 
$u=log(-f)$.  One considers Ricci flows $g=g_t$ with $\gd g\sim \pp_t g=h$ and 
for $({\bf 1B})\,\,\bx^*u=-\pp_tu-\gD u+\dot{{\mc R}}u=0$ (or equivalently $\pp_tf+\gD f-|\na f|^2+\dot{{\mc R}}=0$)
it follows that $\int_Mexp(-f)dV=1$ is preserved and
$\pp_tN={\mf F}$.  Note the Ricci flow equation is $\pp_tg=-2Ric$. 
Extremizing ${\mf F}$ via $\gd{\mf F}\sim\pp_t{\mf F}=0$ involves $Ric+Hess(f)=0$ or
$R_{ij}+\na_i\na_jf=0$ and one knows also that
\bq\label{1.3}
\pp_tN=\int_M(|\na f|^2+\dot{{\mc R}})exp(-f)dV={\mf F};
\end{equation}
$$\pp_t{\mf F}=2\int_M|Ric+Hess(f)|^2exp(-f)dV$$

\section{THE SCHR\"ODINGER EQUATION AND WDW}
\renewcommand{\theequation}{2.\arabic{equation}}
\setcounter{equation}{0}

Now referring to \cite{b9,ccz,zzi,carl,cczz,zzl,c4,c16,zzm,f3,f1,g10,gpi,gbzw,hal,h4,h5,nkl2,
nkl3,pwi,zzu,sa,sb,s35,s36,w3} for details we note first the important observation
in \cite{topp} that ${\mf F}$ is in fact a Fisher information functional.
Fisher information has come up repeatedly in studies of the Schr\"odinger equation
(SE) and the Wheeler-deWitt equation (WDW) and is connected to a differential entropy correspondingto the Nash entropy above (cf. \cite{ccz,carl,g10,gpi}).  
The basic ideas involve 
(using 1-D for simplicity) a quantum potential Q such that $\int_MPQdx \sim{\mf F}$ arising from a wave function $\psi=Rexp(iS/\hbar)$ where $Q=-(\hbar^2/2m)
(\gD R/R)$ and $P\sim|\psi|^2$ is a probability density.  In a WDW context for example one can develop a framework 
\bq\label{2.1}
Q=cP^{-1/2}\pp(GP^{1/2});\,\,\int QP =c\int P^{1/2}\pp(GP^{1/2})
{\mf D}hdx\to
\end{equation}
$$\to -c\int \pp P^{1/2}G\pp P^{1/2}{\mf D}hdx$$
where G is an expression involving the deWitt metric $G_{ijk\ell}(h)$.  In a more
simple minded context consider a SE in 1-D $i\hbar\pp_t\psi=-(\hbar^2/2m)\pp_x^2
\psi+V\psi$ where $\psi=Rexp(iS/\hbar)$ leads to the equations
\bq\label{2.2}
S_t+\frac{1}{2m}S_x^2+Q+V=0;\,\,\pp_tR^2+\frac{1}{m}(R^2S_x)_x=0:\,\,
Q=-\frac{\hbar^2}{2m}\frac{R_{xx}}{R}
\end{equation}
In terms of the exact uncertainty principle of Hall and Reginatto (see \cite
{hal,h5,r3} and cf. also \cite{ccz,zzj,carl,pwi,zzu}) the quantum Hamiltonian
has a Fisher information term $c\int dx(\na P\cdot\na P/2mP)$ added to the classical Hamiltonian
(where $P=R^2\sim|\psi|^2$) and a simple calculation gives
\bq\label{2.3}
\int PQd^3x\sim -\frac{\hbar^2}{8m}\int\left[2\gD P-\frac{1}{P}|\na P|^2\right]d^3x
=\frac{\hbar^2}{8m}\int\frac{1}{P}|\na P|^2d^3x
\end{equation}
In the situation of \eqref{2.1} the analogues to Section 1 involve ($\pp\sim\pp_x$)
\bq\label{2.4}
P\sim e^{-f};\,\,P'\sim P_x\sim -f'e^{-f};\,\,Q\sim e^{f/2}\pp(G\pp e^{-f/2});\,\,
PQ\sim e^{-f/2}\pp(G\pp e^{-f/2});
\end{equation}
$$\int PQ\to -\int \pp e^{-f/2}G\pp e^{-f/2}\sim -\int \pp P^{1/2}G\pp P^{1/2}$$
In the context of the SE in Weyl space developed in \cite{au,av,ccz,carl,c4,
c16,zzm,sa,sb,w3} one has a situation $|\psi|^2\sim R^2\sim P\sim\hat{\rho}=\rho/\sqrt{g}$
with a Weyl vector $\vec{\phi}=-\na log(\hat{\rho})$ and a quantum potential
\bq\label{2.5}
Q\sim -\frac{\hbar^2}{16m}\left[\dot{{\mc R}}+\frac{8}{\sqrt{\hat{\rho}}}
\frac{1}{\sqrt{g}}\pp_i\left(\sqrt{g}g^{ik}\pp_k\sqrt{\hat{\rho}}\right)\right]
=-\frac{\hbar^2}{16m}\left[\dot{{\mc R}}+\frac{8}{\sqrt{\hat{\rho}}}\gD\sqrt{\hat{\rho}}
\right]
\end{equation}
(recall $div grad (U)=\gD U=(1/\sqrt{g})\pp_m(\sqrt{g}g^{mn}\pp_nU)$.  Here 
the Weyl-Ricci curvature is $({\bf 2A})\,\,{\mc R}=\dot{{\mc R}}+{\mc R}_w$
where
\bq\label{2.6}
{\mc R}_w=2|\vec{\phi}|^2-4\na\cdot\vec{\phi}=8\frac{\gD\sqrt{\hat{\rho}}}{\sqrt
{\hat{\rho}}}
\end{equation}
and $Q=-(\hbar^2/16m){\mc R}$.  Note that 
\bq\label{2.66}
-\na\cdot\vec{\phi}\sim-\gD log(\hat{\rho})\sim-\frac{\gD\hat{\rho}}{\hat{\rho}}+
\frac{|\na\hat{\rho}|^2}{\hat{\rho}^2}
\end{equation}
and for $exp(-f)-\hat{\rho}$
\bq\label{2.67}
\int\hat{\rho}\na\cdot\vec{\phi}dV=\int\left[-\gD\hat{\rho}+\frac{|\na\hat{\rho}|^2}
{\hat{\rho}}\right]dV
\end{equation}
with the first term in the last integral vanishing and the second providing 
Fisher information again.
Comparing with Section 1 we have analogues $({\bf 2B})\,\,G\sim (R+|\vec{\phi}|^2)$ with $\vec{\phi}=-\na log(\hat{\rho})\sim \na f$ to go with \eqref{2.4}.
Clearly $\hat{\rho}$ is basically a probability concept with $\int\hat{\rho}dV=1$
and quantum mechanics (QM) (or rather perhaps Bohmian mechanics) seems to enter the picture through the second equation in \eqref{2.2}, namely $({\bf 2C})\,\,\pp_t\hat{\rho}+(1/m)div(\hat{\rho}\na S)=0$ with $p=mv=\na S$, which must
be reconciled with ({\bf 1B}).  In any event the term $G=\dot{{\mc R}}+|\vec{\phi}|^2$
can be written as $({\bf 2D})\,\,\dot{{\mc R}}+{\mc R}_w+(|\vec{\phi}|^2-{\mc R}_w)=
\ga Q+(4\na\cdot\vec{\phi}-|\vec{\phi}|^2)$ which leads to $({\bf 2E})\,\,{\mf F}\sim
\ga\int_MQPdV+\gb\int |\vec{\phi}|^2PdV$ putting Q directly into the picture and
suggesting some sort of quantum mechanical connection.
\\[3mm]\indent
{\bf REMARK 2.1.}
We mention also that Q appears in a fascinating geometrical role in the 
relativistic Bohmian format following \cite{b9,f3,s35,s36} (cf. also \cite{ccz,
carl} for survey material).  Thus e.g. one can define a quantum mass field via
\bq\label{2.7}
{\mf M}^2=m^2exp(Q)\sim m^2(1+Q);\,\,Q\sim \frac{-\hbar^2}{c^2m^2}\frac
{\bx(\sqrt{\rho})}{\sqrt{\rho}}\sim \frac{\ga}{6}{\mc R}_w 
\end{equation}
where $\rho$ refers to an appropriate mass density
and ${\mf M}$ is in fact the Dirac
field $\gb$ in a Weyl-Dirac formulation of Bohmian quantum gravity.  
Further
one can change the 4-D Lorentzian metric via a conformal factor $\gO^2=
{\mf M}^2/m^2$ in the form $\tl{g}_{\mu\nu}=\gO^2g_{\mu\nu}$ and this
suggests possible interest in Ricci flows etc. in conformal Lorentzian spaces
(cf. here also \cite{crwl}).  We refer to \cite{b9,f3} for another fascinating form 
of the quantum potential as a mass generating term and intrinsic self energy.
$\hfill\bs$
\\[3mm]\indent
{\bf NOTE.}
Publication information for items below listed by archive numbers can often be
found on the net listing.$\hfill\bs$

\end{document}